\documentclass[amssymb,aps,twocolumn,showpacs,nofootinbib]{revtex4}
\usepackage[]{graphicx}
\usepackage{hyperref}

\def\ket#1{| #1 \rangle}

\def\bracket#1#2{\langle #1 | #2 \rangle}

\begin{document}

\title{Robust polarization-based quantum key distribution over
  collective-noise channel}

\author{J.-C. Boileau$^1$, D. Gottesman$^2$, R. Laflamme$^{1,2}$, D.
  Poulin$^{1,2}$ and R.W.  Spekkens$^2$}

\affiliation{ $^1$Institute for Quantum Computing, University of
  Waterloo, Waterloo, ON, N2L 3G1, Canada. \\
  $^2$Perimeter Institute for Theoretical Physics, 35 King
  Street N., Waterloo, ON, N2J 2W9, Canada.}

\date{\today}

\begin{abstract}
  We present two polarization-based protocols for quantum key
  distribution. The protocols encode key bits in noiseless
  subspaces or subsystems, and so can function over a quantum
  channel subjected to an arbitrary degree of collective noise, as
  occurs, for instance, due to rotation of polarizations in an
  optical fiber. These protocols can be implemented using only
  entangled photon-pair sources, single-photon rotations,
  and single-photon detectors. Thus, our proposals offer practical and
  realistic alternatives to existing schemes for quantum key
  distribution over optical fibers without resorting to interferometry
  or two-way quantum communication, thereby circumventing,
  respectively, the need for high precision timing and the threat of
  Trojan horse attacks.
\end{abstract}


\maketitle

Quantum key distribution (QKD), such as the BB84 protocol proposed
by Bennett and Brassard in 1984, allows two parties (Alice and
Bob) to generate an arbitrarily long random secret key provided
that they initially share a short secret key and that they have
access to a quantum channel \cite{BB84}. As opposed to classical
key distribution, the secrecy of the generated key does not rely
on computational assumptions but simply on the laws of physics: as
long as quantum mechanics holds, the information available to an
eavesdropper (Eve) can be made arbitrarily small.

Photons are obvious candidates for mediators of quantum
information since they are fast, cheap, and interact weakly with
the environment. Both free air and optical fiber based QKD have
been realized experimentally; see \cite{GRTZ2002} and
\cite{HNDP2002} for reviews.  Any experimental implementation of
QKD naturally has to deal with the issue of noise in the quantum
channel, which substantially complicates the security of QKD, as
Eve may attempt to disguise her eavesdropping as noise
from another source. Standard security proofs deal with channel
noise, including photon loss, and show that Eve acquires
essentially no information provided the noise rate is not too
high.  Higher noise rates mandate lower key generation rates, and
once it becomes too large, secure key generation is
impossible.

Building a viable quantum cryptographic system therefore depends on
ensuring that the noise rate is low. The degree of freedom used to
encode the information can be the polarization of the photon, its
phase, or some combination of both. Purely phase-based schemes have
been realized experimentally \cite{TRT1993} but require complex
interferometric setups, high precision timing, and stable low
temperatures. Interferometry becomes even more challenging with
multi-photon states because of the difficulty of keeping phase
coherence between the photons. A scheme which escapes some of these
limitations using a clever encoding of key bits was proposed
recently~\cite{WASST2003}.

Polarization-based schemes also come with a disadvantage as optical
fibers rotate polarizations of transmitted photons, and the degree of
rotation fluctuates over time.  If left untreated, this would result
in an unacceptably high error rate. A number of proposals have been
made to handle this source of errors; we present a new solution which
is in some ways superior.  Singlet states $\ket{\Psi^-} =
\frac{1}{\sqrt 2}(\ket{01}-\ket{10})$, where $\{\ket{0},\ket{1}\}$ is
any basis of the qubit Hilbert space, have the property that they are
unchanged under equal rotations on both qubits; this is the defining
property of a noiseless subspace. (If one's qubits are the
polarization degrees of freedom of single photons, as we assume here,
then $\ket 0$ and $\ket 1$ can be taken to denote, for instance, the
vertical and horizontal polarization states.) We present two protocols
that take advantage of this property to encode key bits in four- or
three-photon states.  These states should be experimentally
realizable, and form simple examples of a noiseless subspace or
subsystem, respectively (also called a decoherence-free subspace and
subsystem) \cite{ZR1997,KLV2000,KBLW2001}.

Free-space QKD is largely immune to the problem of polarization
rotation: the coupling between the photons and the molecules in
the atmosphere can be absorbed in a dielectric constant to very
good approximation.  The same unfortunately cannot be said about
optical fiber. Rather, the dielectric constant acquires a spatial
and temporal dependence, yielding an overall time dependent
unitary transformation of the polarization state of a single
photon, $U(t)$, as the net effect of the fiber. This varies on the
time-scale of thermal and mechanical fluctuations of the fiber,
the shortest of which we will refer to as $\tau_{fluc}$. If the
time delay between the photons is small compared to $\tau_{fluc}$,
the effect of this noise on the state of $N$ photons is well
approximated by
\begin{equation}
\rho_N \rightarrow [U(t)]^{\otimes N} \rho_N
[U(t)^\dagger]^{\otimes N},
\end{equation}
where $t$ now denotes the time of transmission. This is known as
the unitary collective noise model \cite{ZR1997}.

There are several ways to deal with collective noise. The most obvious
way is to continuously estimate the transformation $U(t)$ and
systematically compensate for it. However, this requires an
interruption of the transmission and, if the fluctuations become too
rapid, the communication channel becomes useless. A second possibility
--- a phase-polarization hybrid which has been used successfully to
realize QKD over $67km$ \cite{SGGRZ2002} --- uses the Faraday
orthoconjugation effect \cite{Martinelli1989} to autocompensate
the effect of $U(t)$. Roughly speaking, if the transformation on
the photon during its transmission from Bob to Alice is described
by $U(t)$, it can be $U(t)^{-1}$ when the photon is transmitted
back from Alice to Bob, yielding no net transformation overall.
The quantum information can be encoded by an extra phase
transformation performed by Alice before returning the photon to
Bob. Obviously, this technique only works if $U(t)$ is roughly
constant throughout the transmission of the photon; this sets an
upper limit of $c\tau_{fluc}$ to the distance over which QKD can
be implemented with this scheme. \footnote{However, with today's
  technology, photon loss is a much more serious limitation to the
  distance over which QKD can be achieved.}

Although such two-way quantum communication can eliminate
collective noise, it allows for new attacks not possible against
BB84. Since Alice receives and emits signals, it is possible for
Eve to probe her laboratory --- a technique known as the {\em
Trojan horse attack}.
There are many ways in which
she can do this.  She could add a weak signal to the channel at a
slightly different frequency and recover some information about
Alice's phase transformation by subsequently filtering the output
signal. Eve could also try to entangle an ancilla system with the
signal before it enters Alice's lab and perform a joint
measurement on the two after Alice has retransmitted the signal.
She could also intercept the signal and send a different signal to
Alice, and thereafter measure the output to estimate the applied
phase transformation, etc. Technical solutions for some of these
attacks have been proposed.  However, Eve has an enormous variety
of attacks available, and to prove true information-theoretic
security, one must assume that Eve has arbitrary technological
power; for instance, she can outperform the best frequency
filtering available to Alice and Bob.  Because of its inherent use
of two-way quantum communication, the protocol is formally quite
different from the standard BB84 protocol, and proving its
security may be quite difficult.%
\footnote{Note, however, that QKD must make use of two-way {\em
classical} communication between Alice and Bob, and this presents
no particular barrier to security proofs.  Indeed, taking full
advantage of two-way classical communication results in a protocol
with substantially greater tolerance for noise \cite{GL2003}.}

The schemes we propose here are purely polarization-based and cope
with collective noise without resorting to two-way quantum
communication.  In the first protocol, the quantum information is
encoded in a noiseless subspace, while in the second protocol, it is
encoded on a noiseless subsystem.\footnote{Such encodings also obviate
  the need for Alice and Bob to share a reference frame (such as, for
  instance, a known relative alignment of their linear polarizers), as
  has been pointed out in the context of quantum communication in Ref.
  \cite{BRS2003}.}  A noiseless subspace is invariant under the
action of the collective noise operation (here $U^{\otimes N}$).  Any
state within it is therefore unaffected (modulo a global unphysical
phase) by the noise. When such a subspace does not exist, it may still
be possible to find a set of density operators which are invariant
under the effect of noise. These density operators instead form a
noiseless subsystem on which pure states can be {\em encoded}.

The protocols we present use singlet states as building blocks.
Information is encoded in the pairing of the photons; the various ways
of organizing three or more photons into pairs provide us with
different states with which to encode information. The photons must be
distinguishable if different pairings are to correspond to different
physical states: they need to be labelled in some way. Physically,
this means that the photons must differ with respect to some degree of
freedom. Here, the photons are assumed to differ in their time of
arrival; they are spatially separated in the optical fiber.
Furthermore, each bit is encoded on multi-photon states which must
also be distinguished even in the presence of noise. Therefore, the
multiplets of photons must be spatially separated by a distance
greater than the separation between individual photons inside a
multiplet. The fluctuation time $\tau_{fluc}$ needs only to be large
with respect to the difference in the arrival times of the first and
last photon of a multiplet.

It is crucial that information about the pairing resides only in the
polarization state of the photon. For example, variations in the
frequency of the photons can reveal pairing information: the
frequencies of the two photons in each singlet must add up to the
frequency of the pump, but the frequencies of photons in different
singlets need not match.  By measuring the energy of the photons, but
not their polarization, Eve can learn about how they are paired
without affecting the outcomes of Bob's measurements.  This energy
signature of the pairing can be eliminated by filtering the photons
before they leave Alice's lab. Indeed, if the band-width of the filter
is smaller than the band-width of the pump laser, almost no
information about the pairing can be recovered by Eve. Similarly, we
must wash out the phase relation between photon pairs. That is, the
information must only be encoded on the {\it relative order} of the
photons, not in their absolute time of arrival. This can be achieved
by choosing the time delay between the photons in each multiplet at
random from a certain range.

We will now present our two protocols: the first one encodes the
quantum information on four-photon states while the second only
requires three-photon states. We do not prove their security here,
but rather point only to their similarities with a protocol
proposed by Bennett in 1992 (B92) \cite{Bennett1992} which is
known to be secure \cite{TKI2002}. We hope to provide a complete
security proof in a later paper. The first protocol requires the
definition of three normalized states of a photon quartet:
\begin{equation}
\begin{picture}(160,60)
\put(5,5) {\makebox(0,0)[c]{$\ket{\psi_3} = $}}
\put(5,25) {\makebox(0,0)[c]{$\ket{\psi_2} = $}}
\put(5,45) {\makebox(0,0)[c]{$\ket{\psi_1} = $}}
\put(130,5) {\makebox(0,0)[c]{$= \frac{1}{\sqrt 2}(\ket a-\ket c)$}}
\put(130,25) {\makebox(0,0)[c]{$= \frac{1}{\sqrt 2}(\ket c -\ket b)$}}
\put(130,45) {\makebox(0,0)[c]{$= \frac{1}{\sqrt 2}(\ket a - \ket b)$}}
\put(30,5) {\circle*{5}}
\put(50,5) {\circle*{5}}
\put(70,5) {\circle*{5}}
\put(90,5) {\circle*{5}}
\put(30,25) {\circle*{5}}
\put(50,25) {\circle*{5}}
\put(70,25) {\circle*{5}}
\put(90,25) {\circle*{5}}
\put(30,45) {\circle*{5}}
\put(50,45) {\circle*{5}}
\put(70,45) {\circle*{5}}
\put(90,45) {\circle*{5}}
\put(30,5) {\line(0,1){12}}
\put(50,5) {\line(0,1){8}}
\put(70,5) {\line(0,1){8}}
\put(90,5) {\line(0,1){12}}
\put(30,25) {\line(0,1){12}}
\put(50,25) {\line(0,1){8}}
\put(70,25) {\line(0,1){12}}
\put(90,25) {\line(0,1){8}}
\put(30,45) {\line(0,1){8}}
\put(50,45) {\line(0,1){8}}
\put(70,45) {\line(0,1){8}}
\put(90,45) {\line(0,1){8}}
\put(30,17) {\line(1,0){60}}
\put(50,13) {\line(1,0){20}}
\put(30,37) {\line(1,0){40}}
\put(50,33) {\line(1,0){40}}
\put(30,53) {\line(1,0){20}}
\put(70,53) {\line(1,0){20}}
\end{picture}
\label{pure_states}
\end{equation}
All these states correspond to pairs of singlet states: in
$\ket{\psi_1}$, photons one and two form a singlet state and so do
photons three and four. The two other states correspond to the two
other possible ways of pairing four photons as is illustrated by
the diagrams.  The states are invariant under uniform rotations,
so in {\em any} basis, these states can be decomposed into the
superpositions noted above, where
\begin{eqnarray}
\ket a &=& \frac{1}{\sqrt 2}(\ket{0101} + \ket{1010}) \nonumber\\
\ket b &=& \frac{1}{\sqrt 2}(\ket{0110} + \ket{1001}) \nonumber\\
\ket c &=& \frac{1}{\sqrt 2}(\ket{0011} + \ket{1100}).\nonumber
\end{eqnarray}

It is straightforward to verify that $|\bracket{\psi_i}{\psi_j}| =
1/2$ for $i \neq j$.  It is therefore impossible to reliably
distinguish any pair of these states, but it is possible to make a
measurement that provides some information.  Measuring the
polarization of all four photons allows Bob to distinguish 
states $\ket{a}$, $\ket{b}$, and $\ket{c}$. Therefore, if Alice
restricts her transmission to one of a pair of states, Bob can
tell which of the two states she sent $50\%$ of the time. For
example, suppose Alice transmits one of the pair
$\{\psi_1,\psi_2\}$. When Bob measures either ``0101'' or
``1010'', he can conclude that she sent $\ket{\psi_1}$.  When he
gets either ``1100'' or ``0011'', he concludes she sent
$\ket{\psi_2}$.  Given any other outcome, Bob can not deduce with
certainty which state she sent.

Note that Bob's measurement constitutes a probabilistic error-free
discrimination procedure \cite{Ivanovic} for any pair of the three
states Alice might submit. That is, it is a measurement that with some
probability identifies, without error, which of the pair was
submitted, and with some probability yields an inconclusive outcome.

We now present the protocol.

\noindent{\em Protocol 1}
\begin{enumerate}
\item Alice chooses a random $(4+\delta)n$ bit string $X$ and a random
  $(4+\delta)n$ trit string $B$.
\item Alice encodes each bit $\{0, 1\}$ of $X$ according to $\{ \psi_1, \psi_2
  \}$ if the corresponding trit of $B$ is $0$; $\{ \psi_2, \psi_3 \}$
  if $B$ is 1; or $\{ \psi_3, \psi_1 \}$ if $B$ is 2.
\item Alice sends the $(4+\delta)n$ quartets of photons to Bob.
\item Bob receives the photons, and announces this fact. For each of
  the $(4+\delta)n$ photon quartets, he randomly chooses between the
  rectilinear or diagonal polarization basis. He then measures each of
  the four photons of each quartet according to this choice of
  basis.
\item Alice announces B. Given this information, and using the
  procedure described above, Bob can determine, for each quartet,
  whether or not his measurement was conclusive, and if conclusive,
  the value of the encoded bit.
\item Alice and Bob discard all bits where Bob's measurement was
  inconclusive. With high probability, there are at least $2n$ bits
  left which they keep. Otherwise, they abort the protocol.
\item Alice selects a random subset of $n$ bits and tells Bob which
  bits were selected.
\item Alice and Bob announce and compare the value of the $n$ selected
  bits to estimate Eve's interference; if more than an acceptable
  number of errors are found, they abort the protocol.
\item Alice and Bob perform information reconciliation and privacy
  amplification on the remaining $n$ bits.
\end{enumerate}

In step 4, the choice of basis does not affect the measurement
outcome of Bob: this is in fact the main property of the encoding.
Nevertheless, it is crucial that Eve does not know in which basis
the measurement is performed.  If she knew, she could measure in
the same basis as Bob, and would know everything Bob knew.  Since
she does not know, she will frequently measure in a different
basis than Bob and therefore introduce errors that will reveal her
presence.

As the protocol is written, in step 6, Alice and Bob discard any
bits for which Bob's measurement is inconclusive.  Nevertheless,
an inconclusive result could still be useful.  Indeed, any
measurement result whose weight differs from $2$, e.g. ``1011'',
indicates that Eve has tampered with the communication. This
provides Alice and Bob with some extra data to estimate Eve's
interference: only allowed code-words should be observed by Bob.

If steps 4 and 5 are inverted --- which could only provide Eve with
more information --- we get a protocol quite similar to B92.  Alice
encodes the bit in two preselected nonorthogonal states which she
sends down the quantum channel. Bob then performs a von Neumann
measurement chosen at random from a certain set, which can be cast in
terms of a positive operator valued measurement (POVM): this is also
required in B92. Nevertheless, there are certain differences in the
nature of these POVMs which must be studied carefully to arrive at a
complete security proof.  We are currently working on these issues.
The B92 protocol is not secure if the transmission rate is below 1/2.
Eve can replace the noisy channel by a perfect channel and measure
Alice's output in such a way that she achieves a conclusive
discrimination with probability ½, in which case, she knows the state
and can send it to Bob. In the case of an inconclusive result, she
does not send anything. From Alice and Bob's point of view, this would
be indistinguishable from the natural noise. By delaying the
announcement of $B$, our proposal escapes this limitation.

The simplicity of the measurement --- single photon polarization ---
is a clear advantage of this protocol. (The possibility of
discriminating between states that are invariant under collective
noise, without having recourse to collective measurements, has also
been noted in Ref.  \cite{Cabello2003}) Furthermore, the required
states have already been produced by several groups
\cite{PDGWZ2001,ZYCZZP2003}. Two singlet states can be produced in a
short time interval via parametric down-conversion by sending a
femto-second pump laser pulse back and forth across a crystal (using a
mirror). Since the photons are emitted in different directions, the
EPR pairs can be clearly distinguished. Zhao et al. \cite{ZYCZZP2003}
have produced these pairs of singlets at a rate of 20 kHz. Optical
delays and switches can be used to create any of the three states of
Eq.  \ref{pure_states}. If Bob's single-photon detectors have a jitter
of roughly 300 picoseconds, the delay lines must have a length of
order 10 centimeters.

At first glance, it looks like there are two copies of the information
in this encoding. For instance, if Alice announces that the bit is
encoded as $\{\psi_1,\psi_2\}$, the value of the first two measurement
outcomes is enough to sometimes deduce the value of the encoded bit:
it is necessarily 1 if the outcomes are the same. The same holds for
the measurement outcomes on the last two photons. Nevertheless, this
redundancy is intrinsic to our quantum encoding scheme and doesn't
provide Eve with any extra information. The second protocol we present
exploits this redundancy to reduce the size of the encoding.

{\em Protocol 2} is a slight modification of {\em Protocol 1}. In step
3, instead of sending the entire state to Bob, Alice randomly discards
one photon from each quartet and sends the remaining three. In step 5,
Alice should also announce which photon she has discarded. Therefore,
the three pure states of Eq.\ref{pure_states} are replaced by the
three mixed states
\begin{equation}
\begin{picture}(100,50)
\put(5,5) {\makebox(0,0)[c]{$\rho_3 = $}}
\put(5,25) {\makebox(0,0)[c]{$\rho_2 = $}}
\put(5,45) {\makebox(0,0)[c]{$\rho_1 = $}}
\put(30,5) {\circle{5}}
\put(50,5) {\circle*{5}}
\put(70,5) {\circle*{5}}
\put(30,25) {\circle*{5}}
\put(50,25) {\circle{5}}
\put(70,25) {\circle*{5}}
\put(30,45) {\circle*{5}}
\put(50,45) {\circle*{5}}
\put(70,45) {\circle{5}}
\put(70,5) {\line(0,1){8}}
\put(50,5) {\line(0,1){8}}
\put(30,25) {\line(0,1){8}}
\put(70,25) {\line(0,1){8}}
\put(50,45) {\line(0,1){8}}
\put(30,45) {\line(0,1){8}}
\put(50,13) {\line(1,0){20}}
\put(30,33) {\line(1,0){40}}
\put(30,53) {\line(1,0){20}}
\end{picture}
\end{equation}
where the ``$\circ$'' denotes the maximally mixed state. These states
are obviously invariant under collective noise.  Furthermore, any pair
can be distinguished with a finite probability, just as with the
states in {\em Protocol 1}. This follows from the fact that they
constitute non-orthogonal mixed states with non-identical supports,
and the fact that one can achieve probabilistic error-free
discrimination of such mixed states \cite{RST2003}. For example,
suppose that Alice has announced that the bit is encoded as
$\{\rho_1,\rho_2\}$. Clearly, any measurement outcome of Bob's where
photon 1 and photon 2 come out parallel rules out the state $\rho_1$.
Here, the two outcomes ``000'' and ``111'' never occur in the absence
of eavesdropping; they indicate that Eve has tampered with the
communication.

These states do not form noiseless subspaces, because any
particular pure state in the decomposition of the density matrices
$\rho_1$, $\rho_2$, or $\rho_3$ does not remain invariant under
collective noise.  Instead it is transformed into another state in
the decomposition of the same density matrix.  For instance,
$\ket{\Psi^-} \otimes \ket{0}$, in the decomposition of $\rho_1$,
becomes under collective bit-flip $\ket{\Psi^-} \otimes \ket{1}$.
The individual states are not noiseless, but the space spanned by
them is; therefore they form a noiseless subsystem, and the
density matrices are invariant under collective noise.

This second protocol shares many similarities with {\em Protocol
1}. It is our hope that essentially the same proof will be able to
show both protocols are secure. In practice, Alice does not have
to create two photon pairs and discard one photon; she could
simply create one pair and one additional photon in the maximally
mixed state. This should greatly increase the optimal transmission
rate since pair creation schemes have relatively low efficiency.
Furthermore, {\em Protocol 2} --- based on trios of photons
instead of quartets --- should suffer less from photon loss and
hence be realizable over greater distances.

On a speculative note, perhaps the two protocols could be hybridized
into a more robust protocol. Alice could always encode her information
on photon quartets. Bob could then divide the photon multiplets into
two sets depending on how many photons from the quartet actually made
it to the destination: set 1 when all four photons made it and set 2
when only 3 photons were detected. The outputs from set 1 could then
be used to complete {\em Protocol 1} while those of the second set
would be used as in {\em Protocol 2}. However, this suggestion
provides Eve with a wide variety of attacks unavailable in our two
protocols, and its security must therefore be studied on a different
basis.

We thank Gilles Brassard for helpful comments on our protocols.
J.-C.B., R.L., D.P., and R.W.S. receive financial support from Canada's
NSERC. This work was partially funded by MITACS and ARDA.

\end{document}